\documentclass[twocolumn,a4paper]{article} 

\newcommand\comment[1]{} 
\newcommand{\sizedef}{
      \headheight=0pt                               
	  \topmargin=-1.5cm \headsep=1.5cm              
      \oddsidemargin=-0.5cm \evensidemargin=-0.5cm  
      \textheight=22truecm \textwidth=16.5truecm    
}
\newtoks\reportnoregister \newtoks\eprintnoregister
\newcommand{\reportnumber}[1]{\reportnoregister={#1}}
\newcommand{\eprintnumber}[1]{\eprintnoregister={#1}}

\reportnumber{\mbox{}} 
\eprintnumber{\mbox{}} 

\newcommand{\reportid}{
   \begin{minipage}{17cm}\vspace{-3.2cm}
     \begin{flushright}
      {\normalsize \the\reportnoregister \\[-.2cm]
	    \eprint{\the\eprintnoregister}}\vspace{3.2cm}
     \end{flushright}
   \end{minipage}\hspace{-17cm} }

\catcode`@=11   
\def\title#1{\gdef\@title{\reportid#1}}
\catcode`@=12   

\newcommand{\eprint}{\textsf} 

\newcommand{\journalfont}{\rm}  
\newcommand{\jou}[1]{{\journalfont #1\ }}
\newcommand{\joudef}[2]{\newcommand #1{\jou{\ignorespaces #2}}}

\joudef{\aaa}    { Astron.\ Astrophys.}
\joudef{\aip}    { Adv.\ Phys.}
\joudef{\adm}    { Adv.\ Math.}
\joudef{\am}     { Ann.\ Math.}
\joudef{\apny}   { Ann.\ Phys.\ (N.Y.)}
\joudef{\apj}    { Astrophys.\ J.}
\joudef{\apjs}   { Astrophys.\ J.\ Suppl.}
\joudef{\cjp}    { Can.\ J.\ Phys.}
\joudef{\cmp}    { Commun.\ Math.\ Phys.}
\joudef{\cqg}    { Class.\ Quantum Grav.}
\joudef{\faa}    { Funct.\ Anal.\ Appl.}
\joudef{\grg}    { Gen.\ Rel.\ Grav.}
\joudef{\ijmpd}  { Int.\ J.\ Mod.\ Phys.\ D}
\joudef{\ijtp}   { Int.\ J.\ Theor.\ Phys.}
\joudef{\invm}   { Invent.\ Math.}
\joudef{\jm}     { J.\ Math.}
\joudef{\jmp}    { J.\ Math.\ Phys.}
\joudef{\jpa}    { J.\ Phys.\ A}
\joudef{\mnras}  { Mon.\ Not.\ R.\ Ast.\ Soc.}
\joudef{\mpla}   { Mod.\ Phys.\ Lett.\ A} 
\joudef{\nature} { Nature}
\joudef{\nc}     { Nuovo Cim.}
\joudef{\npb}    { Nuc.\ Phys.\ B}
\joudef{\ph}     { Physica}
\joudef{\pla}    { Phys.\ Lett. A}
\joudef{\plb}    { Phys.\ Lett. B}
\joudef{\pr}     { Phys.\ Rev.}
\joudef{\prd}    { Phys.\ Rev.\ D}
\joudef{\prep}   { Phys.\ Rep.}
\joudef{\prl}    { Phys.\ Rev.\ Lett.}
\joudef{\prsla}  { Proc.\ Roy.\ Soc.\ Lond.\ A}
\joudef{\ptp}    { Prog.\ Theor.\ Phys.}
\joudef{\ptps}   { Prog.\ Theor.\ Phys.\ Suppl.}
\joudef\rmp      { Rev.\ Mod.\ Phys.}
\joudef\spj      { Sov.\ Phys.\ JETP}

\catcode`@=11

\newcommand\eqalign[1]{\null\,\vcenter{\openup\jot\m@th
  \ialign{\strut\hfil$\displaystyle{##}$&$\displaystyle{{}##}$\hfil
      \crcr#1\crcr}}\,}
\newcommand\meqalign[1]{\null\,\vcenter{\openup\jot\m@th
  \ialign{\strut\hfil$\displaystyle{##}$&&$\displaystyle{{}##}$\hfil
      \crcr#1\crcr}}\,}
\def\ps@reportnumber{%
    \let\@oddfoot\@empty\let\@evenfoot\@empty
    \def\@oddhead{\hfil\rightmark}}
	
\catcode`@=12   

\newdimen\arrayruleHwidth
\makeatletter
\newcommand\Hline{\noalign{\ifnum0=`}\fi\hrule \@height \arrayruleHwidth
  \futurelet \@tempa\@xhline}
\makeatother

\newcommand\thickbaselines{\baselineskip=20pt\lineskip=3pt\lineskiplimit=3pt}

\catcode`@=11

\renewcommand\matrix[1]{\null\,\vcenter{\thickbaselines\m@th
    \ialign{\hfil$##$\hfil&&\quad\hfil$##$\hfil\crcr
      \mathstrut\crcr\noalign{\kern-\baselineskip}
      #1\crcr\mathstrut\crcr\noalign{\kern-\baselineskip}}}\,} 
\catcode`@=12   

\newcommand\be{\begin{equation}} \newcommand\ee{\end{equation}} 
\newcommand\bd{\begin{displaymath}}\newcommand\ed{\end{displaymath}}

\newcommand\undersim[1]{\mathop{\vtop{\ialign{##\crcr
     $\hfil\displaystyle{#1}\hfil$\crcr\noalign
     {\kern1pt\nointerlineskip}\hbox{$\hfil\sim\hfil$}\crcr
     \noalign{\kern1pt}}}}}

\newcommand{\smallcaption}[1]{\caption{\protect\small#1}}

\newcommand{\acronym}[3]{\newcommand{#1}{#3 (#2)\relax\renewcommand{#1}{#2}}}

\newcommand\eg{{\it e.g.}}

\sizedef

\usepackage{amsmath}
\usepackage{graphicx}

\usepackage{amssymb}

\acronym{\hw}{HW}{{Harrison-Wheeler}}
\acronym{\gbone}{GB1}{{\em generalized Buchdahl $n=1$ polytrope}}
\acronym{\gbfive}{GB5}{{\em generalized Buchdahl $n=5$ polytrope}}
\acronym{\SSS}{SSS}{{\em static spherically symmetric}}
\acronym{\TOV}{TOV}{{\em Tolman-Oppenheimer-Volkoff}}
\acronym{\NU}{NU}{{\em Nilsson-Uggla}}

\begin{document}

\reportnumber{USITP 2000-13}

\title{\Large Ultracompact stars with multiple necks}
           \author{ Max Karlovini\footnote{E-mail: \eprint{max@physto.se}}\;, 
     Kjell Rosquist\footnote{E-mail: \eprint{kr@physto.se}}\; and 
     Lars Samuelsson\footnote{E-mail: \eprint{larsam@physto.se}} \\[10pt]
        {\small Department of Physics, Stockholm University}  \\[-10pt]
        {\small Box 6730, 113 85 Stockholm, Sweden} \\
\begin{minipage}[t]{0.8\linewidth}\small{ We discuss ultracompact stellar objects
which have multiple necks in their optical geometry.  There
are in fact physically reasonable equations of state for which the number of
necks can be arbitrarily large.  The proofs of these statements rely on a
recent regularized formulation of the field equations for static spherically
symmetric models due to Nilsson and Uggla.  We discuss in particular the
equation of state $p= \rho- \rho_{\mathrm{s}}$ which plays a central role in
this context.   }\end{minipage}}

\date{}

\maketitle



The purpose of this communication is to discuss the existence of 
ultracompact relativistic stellar models which have multiple necks in their 
optical geometries.  A few examples of such models were given in
\cite{krs:annalen}. Abramowicz and coworkers discussed the optical geometry 
of uniform density stars in \cite{aabgs:optical}.  Such stars have a single 
neck and an accompanying bulge.  Necks (and bulges) correspond to unstable 
(stable) circular null orbits.  We say that a star is ultracompact if the 
spacetime admits one or more spatially closed null orbits.  This definition of 
ultracompactness was proposed in \cite{rosquist:trapped}.  As is well-known, a 
Schwarzschild black hole has exactly one closed null orbit at $r=3M$.  A 
uniform density star is ultracompact if its radius satisfies $R\leq 3M$.  
However, for more realistic equations of state, there is no such simple 
relation between radius and ultracompactness \cite{rosquist:trapped}.  In the 
past, ultracompactness was usually defined by the condition $R\leq 3M$.  
However, as was shown in \cite{rosquist:trapped} that definition does not 
capture what may be regarded as the essence of ultracompact bodies, namely 
their ability to serve as a gravitational trap for relativistic matter such as 
neutrinos, electromagnetic waves or gravitational waves.  In particular, the 
neck (or necks) of a compact star can be located entirely within the stellar 
interior.

The optical geometry is a very useful tool to visualize the structure of the
null geodesics of \SSS\ models.  Models with $R<3M$ have a neck in the
exterior optical geometry which is associated with the Schwarzschild circular
null orbit at $R=3M$.  In the interior of the star there are also other
spatially closed null orbits.  This was believed to be the typical situation
for ultracompact stars.  However, as will be discussed below, the optical
geometries for a large class of more realistic equations of state have
multiple necks.  The maximum number of necks, $n_{\rm max}$, for a given
equation of state is sensitive to the number $\gamma_\infty= 1+
\lim_{p\rightarrow \infty}p/\rho$ (assuming that it exists).  If the high
density equation of state has the Zel'dovich stiff matter form, $p=\rho$, so
that $\gamma_\infty =2$, then the number of necks becomes arbitrarily large as
the central density, $\rho_\mathrm{c}$, tends to inifinity.  On the other hand
consider the equation of state $p = (\gamma-1) (\rho-\rho_{\rm s})$ where
$\gamma (= \gamma_\infty)$ and $\rho_\mathrm{s}$ are constants.  In the limit
$\gamma \rightarrow \infty$ it goes over into the uniform density equation of
state for which we know that $n_{\rm max}=1$.  Thus $n_{\rm max}$ varies in
the range $[1,\infty)$ as $\gamma$ varies from infinity to 2.  At the other
end $n_{\rm max}=0$ for $\gamma$ below a certain value $\gamma_0>4/3$
\cite{cve:photsurf}.  These results and others mentioned in this communication
will be proved and discussed in more detail in \cite{krs:multneck}.

The above statements can be inferred from a powerful formulation of the \SSS\
field equations due to Nilsson and Uggla  \cite{nu:grstars_linear,
nu:grstars_polytropic}.  The idea is to
replace the \TOV\ equations by a maximally reduced dynamical system which has
a state space which is both regular and compact. In fact Nilsson and
Uggla derives three such systems, one adapted to linear equations of state
\cite{nu:grstars_linear}, and two complementary for polytropic
equations of state \cite{nu:grstars_polytropic}. We will use the
$\Sigma$-$K$-$y$ system
given in \cite{nu:grstars_polytropic}, to be referred to as the
\emph{loaf of bread}  
system and the loaf of bread state space due to the shape of the state 
space.  A price one has to pay is that the system
is necessarily 3-dimensional and so has one extra dimension compared to the
\TOV\ system.  As a result, the loaf of bread system incorporates models which are
irregular as well as regular at the center.  In fact the inclusion of the
irregular models is an essential feature in order to understand the structure
of the solution space of the regular models.  The underlying reason for this
will be touched upon below.  For simplicity, in this paper we focus attention
on the stiff fluid equation of state $p= \rho -\rho_{\rm s}$ where $\rho_{\rm
s} \neq0$ and the subindex ``s'' stands for evaluation at the surface of the
star where $p=0$.  This is the most general equation of state for which the
speed of sound (defined as $\sqrt{dp/d\rho}$) equals the speed of light.  In
particular, there is no manifest violation of causality in contrast to the
case of uniform density matter.


We now proceed to show how the loaf of bread dynamical system and state space for 
\SSS\ stars can be used to analyse compactness characteristics such as the 
occurrence of spatially closed null orbits.  We begin by introducing the state 
space.  The \SSS\ metric is written in the form
\begin{equation}
   ds^2 = -e^{2\nu}d t^2 + N^2 d x^2
                         + r^2(d\theta^2+\sin^2\!\theta\,d\phi^2)\ ,
\end{equation}
where $\nu$, $N$ and $r$ are functions of the radial variable $x$.  The 
function $N$ is the radial gauge and below we will make a definite gauge 
choice which simplifies the field equations.  The radial gauge function can be 
viewed as the analogue of the lapse function for spatially homogeneous models.  
Also, analogously, the expansion of the unit normals to the homogeneous 
hypersurfaces plays an important role in obtaining a compact state space.  See 
\cite{ujr:hh} and references therein for a discussion of the lapse versus 
radial gauge and the use of the expansion for creating a compact state space.  
An expansion related function is given by
\begin{equation}
        \Theta = \sqrt3 N^{-1}\frac{d}{d x} (\nu+\ln r) \ .
\end{equation}
Next we introduce two dimensionless gravitational variables according to
\begin{equation}
 \Sigma := \sqrt3 N^{-1} \Theta^{-1} \frac{d \nu}{d x} \ ,\qquad
      K := 3 r^{-2} \Theta^{-2} \ .
\end{equation}
A third dimensionless variable is defined as $y:= p/ (\rho+p)$.  We shall also 
need the matter function
\begin{equation}
        F(y) := (\rho+p)\frac{d y}{d p} = 1-y-y v^{-2}
              = 1-y-\gamma^{-1} \ ,
\end{equation}
where $v^2 = dp/d\rho$, is the speed of sound and $\gamma := 
y^{-1}v^2$ is the relativistic compressibility index.  Now choosing $N = \sqrt3 y 
\Theta^{-1}$ and using a prime to denote differentiation with respect to $x$, 
the field equations reduce to the dimensionless system
\begin{equation}\label{eq:reduced}
 \begin{split}
    \Sigma' &= -yK\Sigma + \tfrac12 P[1+2y(1-2\Sigma)] \ ,\\
            K' &= 2y(1-K-2P)K \ ,\\
            y' &= -yF(y)\Sigma \ ,
\end{split}\end{equation}
where $P:= 1-\Sigma^2-K$, together with the decoupled equation for $\Theta$ 
given by 
\begin{equation}
        \Theta' = -y(K-\Sigma+2\Sigma^2) \Theta \ .
\end{equation}
The state space is defined by the inequalities $|\Sigma| \leq1$, $P\geq0$ and 
$0\leq y\leq y_1 \leq 1$ where $y_1$ is defined to be the lowest
positive value
of $y$ satisfying $F(y)=0$.  The state space is 
a compact region having the form of a half pipe resembling a loaf of bread 
\cite{nu:grstars_polytropic}.  The bottom lies in the surface $K=0$ and the ``roof 
ridge'' stretches along the line $\{\Sigma=0, K=1\}$.  The surface of 
the star is located on the curve defined by $y=P=0$ while the center 
corresponds to the points of the roof ridge \cite{nu:grstars_polytropic}.  For the 
stiff fluid equation of state, $y = p/(2p+ \rho_{\rm s})$, implying that $y$ 
varies in the range $0\leq y<y_1=1/2$ where $y_1=y_\infty := 
\lim_{p\rightarrow \infty}y$.  Further $F(y)= 1-2y$ so $F$ tends to zero only 
in the limit of infinite pressure.  The entire state space is therefore 
completely specified in this case by the above range of $y$.  Also, the system 
\eqref{eq:reduced} is polynomial for this equation of state and hence
regular throughout the state space.  
Note, however, that this nice and simple state of affairs does not always hold 
for more general equations of state for which $y$ is not a monotone function 
of $p$. The function $F(y)$ then has zeroes for values of $y$
corresponding to finite pressures. This situation occurs \eg { }for the
Harrison-Wheeler equation of  state. A more general formulation valid
for a much larger class of equations of state will be given in \cite{krs:multneck}.

The compactness of an object has usually been measured by the single number 
$R/M$, {\em i.e.}, the tenuity in the terminology of \cite{rosquist:trapped}.  
In the loaf of bread system the tenuity is simply given by
\begin{equation}\label{eq:tenuity}
        R/M = 1+1/\Sigma_{\mathrm{s}} \ .  
\end{equation}
However, as discussed above and in \cite{rosquist:trapped}, the tenuity by 
itself does not give the whole story for general equations of state.  Instead 
the key to a more satisfactory understanding of stellar compactness is the 
effective potential which governs the null geodesics.  It is also known as the 
centrifugal potential and is given by \cite{cf:gwresonance}
\begin{equation}
        V = l(l+1) e^{2\nu}r^{-2} \ .
\end{equation}
A star is ultracompact precisely if $V$ has an extremum for some $r>0$.  The 
evolution equation for $V$ takes the simple form
\begin{equation}\label{eq:Vevol}
        V' = -2y(1-2\Sigma) V \ .
\end{equation}
It follows that the extrema of $V$ are located in the plane $\Sigma= 1/2$.  
Note that while the right hand side of \eqref{eq:Vevol} also vanishes for 
$y=0$ this is just a mathematical artifact caused by the vanishing of the 
chosen radial gauge at the surface.  Of course, $V$ may have an extremum at 
the surface but only if $\Sigma= 1/2$ there.

\begin{figure}[!tbp]
 \centering
 \begin{minipage}[t]{0.8\linewidth}
  \centering\includegraphics[width=\textwidth]{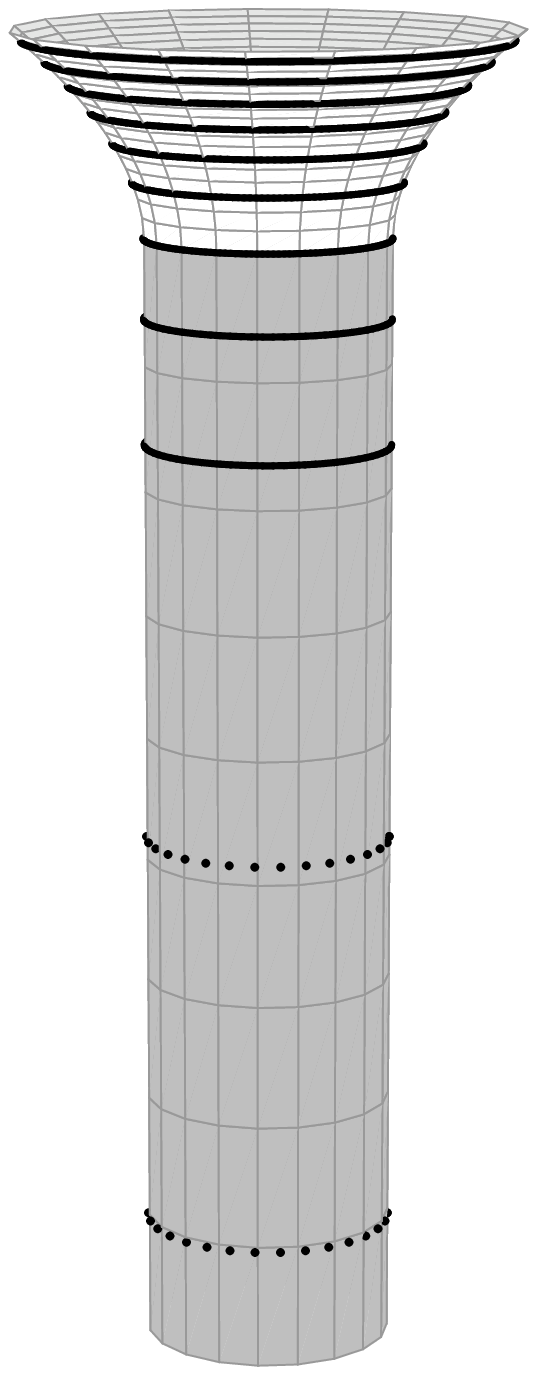}
                        \smallcaption{The equatorial submanifold of the optical geometry 
                        of the skeleton solution for the stiff matter equation of state 
                        ($p=\rho-\rho_\mathrm{s}$).  The figure represents the star from 
                        $r=0.005\,M$ to $r=9.5\,M$.  The cylinder is in reality infinitely 
                        long in the limit $r \rightarrow 0$.  The fat circles represent 
                        the surfaces $r= 1,2,\ldots,9$ while the dotted curves represent 
                        the surfaces $r=0.1$ and $r=0.01$.  The shaded region represents 
                        the stellar interior.}
        \label{fig:stiff_skeleton}
   \end{minipage}
\end{figure}

For any given equation of state there is a special ``skeleton solution'' 
representing the extreme case of infinite pressure at the center.  
Although unphysical, the skeleton solution is very important for 
understanding the global properties of the solution space of \SSS\ 
systems.  In particular, the state space orbit of a typical relativistic 
configuration, after starting out from the roof ridge near $y=y_\infty$, 
spirals around the skeleton orbit before finishing at the surface, 
$y=0$.  It so happens that the skeleton solution for the stiff fluid can 
be given in the exact form
\begin{equation}\label{eq:skeleton}
        \Sigma = \tfrac12 \ ,\qquad K = \frac3{4(1+y)} \ .
\end{equation}
This is a special Tolman type V solution \cite{tolman:sssfluids}.  Thus the 
skeleton solution has $\dot{V}=0$ and hence $V= constant$ throughout its interior.  
It follows that there is a continuous family of circular null geodesics 
extending from the center to the surface of this stellar model.  Now using 
\eqref{eq:tenuity} and \eqref{eq:skeleton} it follows that $R/M=3$ for the 
skeleton solution.  The model can be conveniently visualized by its optical 
geometry \cite{aabgs:optical} which is shown in figure 
\ref{fig:stiff_skeleton}.
\begin{figure}[!tb]
 \centering \begin{minipage}[t]{0.8\linewidth}
  \centering
  \includegraphics[width=\textwidth]{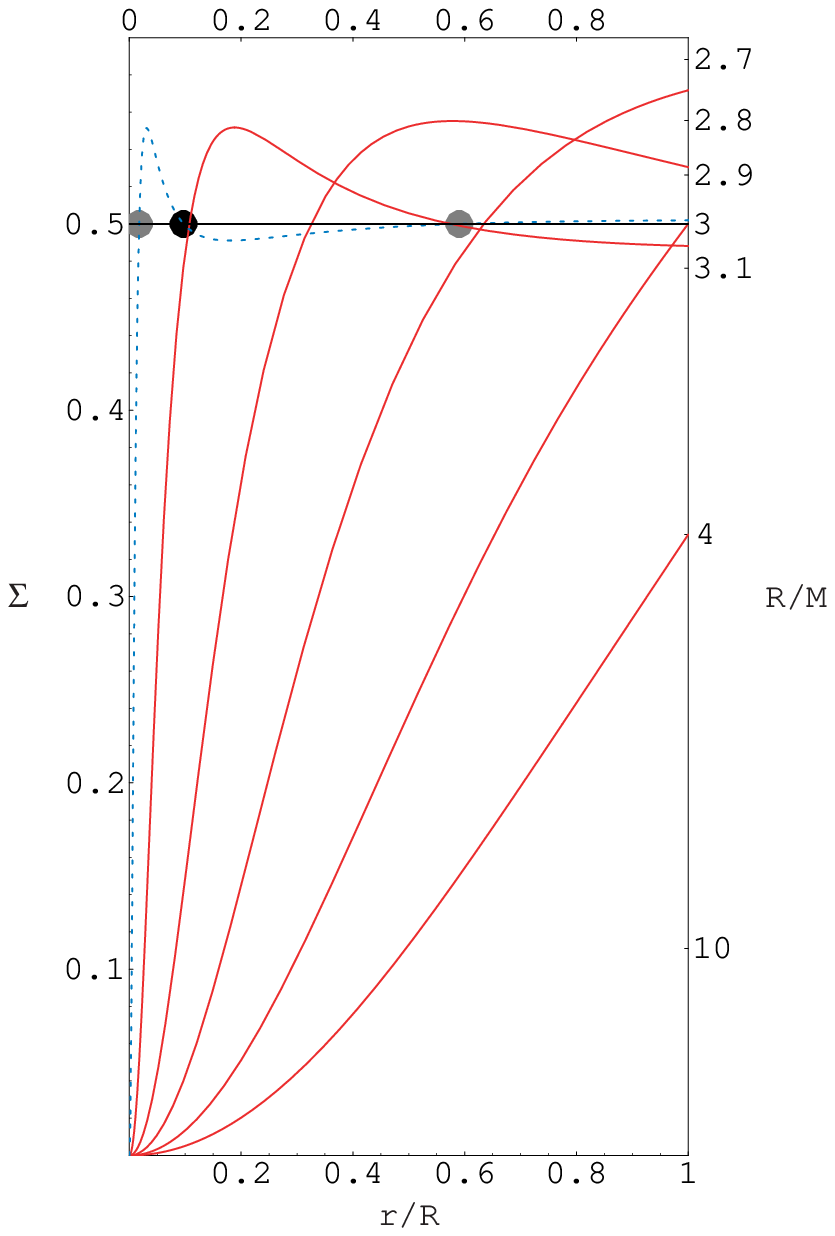}
                        \smallcaption{Plot of $\Sigma$ versus Schwarzschild radial 
                        coordinate for some selected values of the central density.  The 
                        horizontal line at $\Sigma=\frac12$ represents the skeleton 
                        solution.  The remaining curves correspond to the central 
                        densities (using clockwise enumeration about the upper left corner 
                        of the plot), $\rho_{\mathrm{c}}/\rho_{\mathrm{s}} = 1.329, 2.09, 
                        6.2, 30, 274$ and $9700$.  The most tenuous of the plotted models 
                        has $R/M=4$ while the next has $R/M=3$ and is the first 
                        ultracompact model in the sequence.  The third model has the 
                        smallest tenuity in the sequence, $R/M=2.75$.  It is a single neck 
                        model like the fourth and fifth models.  The sixth model is a 
                        double neck model represented by the dotted curve.  Its inner neck 
                       is indicated by the circular disk while the positions of its two 
                        bulges are given by the circles.  Its outer neck is the 
                        Schwarzschild exterior neck at $r=3M$.  }
 \label{fig:stiff_sigma}
 \end{minipage}
\end{figure}

To understand the solution space it is necessary to analyze the full 
3-dimensional system \eqref{eq:reduced}.  However, for simplicity, in this 
communication we restrict attention to the behavior of $\Sigma$.  The necks 
and bulges can then be read off as the zeros of the function $\Sigma-\frac12$.  
Necks and bulges correspond to unstable and stable circular null orbits 
respectively.  The detailed geodesic structure requires a more careful 
analysis which is beyond the scope of this communication.

The crucial observation comes from the realization that the orbits which 
correspond to sufficiently compact objects will spiral about the skeleton 
solution.  This happens because the skeleton solution corresponds to a 
separatrix orbit going from a spiral critical point on the state space 
boundary at $\Sigma=\frac12$, $K=y=\frac12$.  Since the skeleton solution 
sits on the surface $\Sigma=\frac12$ the number of necks is determined by the 
number of revolutions in the $\Sigma$-$K$ plane. In the limit of infinite 
central pressure the number of revolutions grows without limit and hence the 
number of necks can be arbitrarily large.  In figure \ref{fig:stiff_sigma}, 
$\Sigma$ is plotted as a function of the Schwarzschild variable $r$ for some 
values of the central density including a value which leads to a double neck 
star.

The null geodesic structure is intimately connected with quasi-normal 
perturbation modes of the relativistic stellar models 
\cite{cf:gwresonance,aabgs:optical}.  These modes in turn are closely related 
to gravitational wave scattering.  Threfore the appearance of multiple necks 
is likely to influence the picture of both gravitational pulsation modes and 
gravitational wave scattering.  The examples of multiple neck models we have 
given in this communication all correspond to unstable stellar objects.  
Although this would seem to limit the physical applicability of the results, 
moderately unstable models may in fact have a role to play as intermediate 
stages in gravitational collapse situations \cite{nc:criticalperfect}.  It is 
an open problem whether there exist multiple neck models which are both stable 
and causal. We mention without proof that the stiff fluid equation of state 
admits a sequence of stable (and of course causal) single neck models but no 
stable double or higher neck models.

\subsection*{Acknowledgements}

We would like to thank Ulf Nilsson and Claes Uggla for giving us access to
their results at an early stage prior to publication and for numerous useful
discussions.  Many thanks are also due to Ingemar Bengtsson, Martin Goliath 
and S\"oren Holst for their interest and many valuable comments
regarding this work. Financial support was given by the
Swedish Natural Science Research Council.


\bibliographystyle{prsty}
\bibliography{kr}

\end{document}